\documentstyle[12pt]{article}
\renewcommand{\baselinestretch}{1.4}
\begin{document}
\title{Chiral approach to weak radiative hyperon decays and the
$\Xi ^0 \rightarrow \Lambda \gamma $ asymmetry}
\author{
{P. \.{Z}enczykowski}$^*$\\
\\
{\em Dept. of Theor. Physics} \\
{\em Institute of Nuclear Physics}\\
{\em Radzikowskiego 152,
31-342 Krak\'ow, Poland}\\
}
\maketitle
\begin{abstract} We reanalyse the recent version of 
the chiral model of weak radiative hyperon decays, 
proposed by Borasoy and Holstein.
It is shown that 
predictions of the analysed model are significantly changed when
one accepts the usual classification of
$\Lambda (1405)$ as an $SU(3)$-singlet. 
In particular, for the $\Xi ^0 \rightarrow \Lambda \gamma $ decay 
large negative asymmetry is obtained. 
This is contrasted with the Hara's-theorem-violating approaches where this
asymmetry is large and positive.
\end{abstract}
$^*$ E-mail:zenczyko@solaris.ifj.edu.pl\\
\vskip 0.8in
\begin{center}REPORT \# 1833/PH, INP-Krak\'ow \end{center}
\newpage

\section{Introduction}
In 1964 Hara proved a theorem \cite{Hara}, according to which the 
parity-violating amplitude of the $\Sigma ^+ \rightarrow p \gamma $
decay should vanish in the limit of exact $SU(3)$ symmetry.
For weak breaking of $SU(3)$, one then expects 
a small asymmetry in this decay.
The experimental evidence accummulated over the years proves, however, that 
the asymmetry in question is large and negative \cite{Foucher,LZ},
$\alpha (\Sigma ^+ \rightarrow p \gamma ) = -0.76 \pm 0.08$. 
Understanding this experimental result and related data
on other weak radiative hyperon decays (WRHD's) constitutes
an important issue for low-energy physics of weak interactions.

WRHD's were studied in many approaches (for a review see ref.\cite{LZ}). 
Generally, in most models in which Hara's theorem is satisfied,
a small value of the 
$\alpha (\Sigma ^+ \rightarrow p \gamma )$ asymmetry
is predicted (ref.\cite{Orsay} is an important exception here).
This was in particular the case of the chiral approach
 in which it was found \cite{Neufeld} that 
$|\alpha (\Sigma ^+ \rightarrow p \gamma )|<0.2 $ .
Recently, Borasoy and Holstein (BH) attempted a new description of WRHD's
within the chiral framework \cite{Holstein99}.  
In the BH approach, pole model
contributions from low-lying excited $J^P=1/2^+$ intermediate states 
were studied, in addition to the usually considered contributions from
the ground-state baryons and the $1/2^-$ baryon resonances.  
Model parameters were determined from a fit to nonleptonic
hyperon decays and used as an input for the description of WRHD's.
It was found that in the parity-conserving amplitudes,
the contribution of the $1/2^+$ resonances 
is substantial (especially for the $\Sigma ^+ \rightarrow p \gamma $ decay).
Furthermore, a large negative $\Sigma ^+ \rightarrow p \gamma $
asymmetry (around $-0.50$) was obtained. 
Although the detailed BH predictions do not fit the WRHD data well, the
observation
that inclusion of the $1/2^+$ resonances permits
a significant contribution
to the parity-conserving  $\Sigma ^+ \rightarrow p \gamma $ amplitude 
is interesting.
With the inclusion of the $1/2^+$ resonances,
the parity-conserving $\Sigma ^+ \rightarrow p \gamma $ amplitude does not
vanish in the $SU(3)$ limit.
Such vanishing, occuring when only ground-state baryons
are considered as intermediate states,
constituted a problem for the authors of ref.\cite{Orsay}.
In their paper, 
the parity-conserving $\Sigma ^+ \rightarrow p \gamma $ amplitude 
depended on the difference (dissappearing in the $SU(3)$-limit) 
of the anomalous parts of the $\Sigma ^+$ and 
$p$ magnetic moments.
Predictions published in \cite{Orsay} were obtained using experimental values
of the relevant magnetic moments.
On the other hand, if quark model formulas for $\mu _{\Sigma ^+}, \mu _p$ 
are employed, these predictions lead to a positive sign for the 
 $\alpha (\Sigma ^+ \rightarrow p \gamma )$ asymmetry \cite{Orsay}.
 Thus, it is certainly interesting that 
 a contribution to parity-conserving amplitudes,
 which does not vanish in the $SU(3)$ limit 
  and leads to a definitely negative asymmetry, was identified
 in ref.\cite{Holstein99}.

Although in the BH paper 
the $\Sigma ^+ \rightarrow p \gamma $ asymmetry is fairly large
(this was also the case in ref.\cite{Orsay}),
other BH predictions do not seem to be in good
agreement with experiment. 
One such prediction is the branching
ratio of the decay $\Sigma ^+ \rightarrow p \gamma $, which 
is larger than data by an order of magnitude.
The huge size of this branching ratio
stems directly from
the large value of the relevant parity-conserving amplitude.
(In ref.\cite{Orsay} this branching ratio, although somewhat smaller 
than the experimental one, is still of reasonable size.)
This discrepancy suggests that
the size of contributions from excited $1/2^+$ states
is overestimated in ref.\cite{Holstein99}.
Another problem for ref.\cite{Holstein99} is that, in those places where
the contribution from the $1/2^+$ excited 
states is already small, there is no
agreement between the predictions of ref.\cite{Holstein99} and 
the original model of ref.\cite{Orsay}.  
It seems natural that such agreement should exist because 
both Gavela et al. \cite{Orsay} as well as Borasoy and Holstein 
\cite{Holstein99} intended
to include all important contributions from the
ground-state and excited $1/2^-$ baryons in their calculations.
Thus, 
there must be an additional difference between the two approaches.
In the present paper we analyze where this difference comes from and
show its main reason. 
It turns out that {\em in fact} (and contrary to the statements 
contained therein)
ref.\cite{Holstein99} omits contributions from 
the low-lying $J^P=1/2^-$ excited {\em SU(3)-singlet} baryon $\Lambda (1405)$. 
When this singlet contribution is taken into account, one
reproduces more or less closely the predictions of ref.\cite{Orsay}.
In particular, the chiral BH approach with the contribution 
of intermediate singlet baryon included
leads to a significantly negative asymmetry 
in the $\Xi ^0 \rightarrow \Lambda \gamma $ decay.

\section{Intermediate states in parity-violating amplitudes}

Prescriptions of the chiral approach of ref.\cite{Holstein99} reduce 
to the familiar formulas of the pole model. Thus, what is analysed in ref.
\cite{Holstein99} is a pole model consistent with chiral symmetry conditions.
It is this pole model that is ultimately compared with experiment.
Clearly, predictions of any pole model depend in an essential way
on the intermediate states taken into account. 
As the intermediate states ($B^*$ in Fig. 1) in the parity-violating amplitudes,
the authors of ref. \cite{Orsay} consider {\em all} those $J^P=1/2^-$
states from the $(70,1^-)$ multiplet
which have appropriate flavour quantum numbers.
That is, they consider all relevant states from the
two $SU(3)$ octets $^2{\bf 8}$ and 
$^4{\bf 8}$ (of quark spins 
$S=1/2$ and $S=3/2$) and a singlet $^2{\bf 1}$ (of quark spin $S=1/2$).  
(Contributions from the decuplet $^2{\bf 10}$ vanish.)
In the BH paper, on the other side,
only states from the 
lowest-lying octet of the excited $J^P=1/2^-$ baryons 
($^2{\bf 8}$ in theoretical models) are taken into account. 
While one may perhaps expect that contributions from the 
higher-lying octet $^4{\bf 8}$ are not very important, 
neglecting the singlet
is not justified as explained below.

In ref.\cite{Holstein99} the $\Lambda (1405)$ baryon is treated
as a member of the lowest-lying octet of excited $J^P=1/2^-$ baryons.
However,
$\Lambda (1405)$ is usually classified as a singlet \cite{PDG},
while the octet $\Lambda $ is identified with $\Lambda (1670)$.
A corresponding $\Sigma $ state is found at 1750 MeV.
The PDG flavour assignment of the lowest-lying $J^P=1/2^-$ states \cite{PDG}
is corroborated by theoretical calculations and decay analyses \cite{LCH}.
Isgur and Karl \cite{IK78} predict a dominantly singlet state
just below 1500 MeV, and a dominantly octet $\Lambda $ state at around 1650 MeV.
Large spin-orbit 
splitting between the $\Lambda (1405)$ and the $J^P=3/2^-$ state
$\Lambda (1520)$ (also a singlet) is not reproduced in their model, though. 
This discrepancy
between the predictions of quark model and experiment, as well as 
proximity of $\Lambda (1405)$ mass to the sum of $N$ and $\bar{K}$ masses
are sometimes regarded as 
an indication that $\Lambda (1405)$ is an unstable $N\bar{K}$ bound state 
akin to the deuteron, and
not a quark-model $SU(3)$-singlet resonance. 
However, as stressed by Dalitz in ref. \cite{PDG},  
in order to accommodate the quark-model singlet $J^P=1/2^-$ state,
another $\Lambda$ baryon
should then be present in the vicinity of $\Lambda (1520)$.  
No such state has been observed experimentally.
Furthermore, it turns out that inclusion of hadron-loop effects 
(i.e. of the coupling to the $N\bar{K}$ channel in particular) 
splits
the two $J^P = 1/2^-$ and $J^P=3/2^-$ quark-model  $SU(3)$-singlet 
 $\Lambda $ states in the correct way 
with the $J^P=1/2^-$ state shifted down in mass \cite {TZ86,Zen86}.
All other hadron-loop-induced shifts and mixings in the whole $(70,1^-)$
multiplet (as well as those of ground-state baryons \cite{TZ84}) 
are also in good agreement
with the data.
In particular, in the model of ref.\cite{TZ86}
the downward shift of $\Lambda (J^P=1/2^-)$ 
is substantial (though somewhat small).
All this shows clearly that $\Lambda (1405)$ should indeed be considered an
approximate SU(3)-singlet, and not an octet.
Thus, when taking into account the lowest-lying $J^P=1/2^-$ states,
in addition to the $N(1535)$, the $\Sigma (1750)$, and the $\Xi (?)$,
we have to include {\em two} $\Lambda $ states: the dominantly 
singlet $\Lambda (1405)$ and the dominantly octet $\Lambda (1670) $.

\section{Relative size of singlet and octet contributions}

Clearly, SU(3) symmetry cannot predict the relative size and sign of 
the contributions from
the singlet and octet excited $\Lambda $'s.
To get this crucial information, one has to employ 
a broader symmetry that would
put all the considered $J^P=1/2^-$ states into a single multiplet.
This is usually achieved through the use of $SU(6)\times O(3)$ 
symmetry of the quark model.
Such an approach was employed in ref.\cite{Orsay}.
In order to see how important the
contributions neglected in the BH paper are,
we must therefore study ref.\cite{Orsay} 
in more detail.

With the help of Tables 1 and 2 of ref.\cite{Orsay}, 
the contributions from various intermediate states (i.e. from each of the
two octets and from the singlet) may be easily reconstructed.
These contributions are gathered in Table 1 here.  Normalization of entries
in Table 1 is such that the totals for each decay are equal to the numbers
given in Table 7.2 of ref.\cite{LZ} multiplied by 
a common factor $2+K$. 
This reflects full agreement between the $SU(6)_W \times VMD$ approach of
\cite{Zen89,Zen91} (which in turn is based on ref.\cite{DDH})
and the approach of ref.\cite{Orsay} (apart from the question of the
relative sign of contributions from diagrams $(b1)$ and $(b2)$ in Fig.1).
The parameter $K$ is of order $1$.  
In ref.\cite{Orsay} it is calculated within the framework of the 
harmonic oscillator constituent quark model as 
$K=R^2\omega ^2 = \omega/m = 
500~MeV/400~MeV =1.25$ with $R$ being baryon radius, $\omega$ - excitation
frequency and $m$ - constituent quark mass.
\\

Table 1

Weights of amplitudes corresponding to diagrams $(b1)$ and $(b2)$ in ref.
\cite{Orsay}.

\renewcommand{\baselinestretch}{2.0}
\begin{center}
\begin{tabular}{|l|c|c|c|}
\hline
process & $^{2s+1}{\bf R}_{SU(3)}$ & diagram (b1) & diagram (b2) \\ [0.3 cm]
\hline
$\Sigma ^+ \rightarrow p \gamma $ 
& $^2{\bf 8}$ 
& $-\frac{1}{3\sqrt{2}}(2+K)$&$-\frac{1}{3\sqrt{2}}(2+K)$\\ [0.1 cm]
&$^4{\bf 8}$
&0&0\\ [0.1 cm]
\hline
&total&$-\frac{1}{3\sqrt{2}}(2+K)$&$-\frac{1}{3\sqrt{2}}(2+K)$\\ [0.3 cm]
\hline
\hline
$\Lambda \rightarrow n \gamma $
&$^2{\bf 8}$ 
&$\frac{1}{6\sqrt{3}}(2+\frac{K}{3})$&$\frac{1}{3\sqrt{3}}(2+\frac{K}{3})$
\\ [0.1 cm]
&$^4{\bf 8}$&$\frac{1}{9\sqrt{3}}K$&$\frac{2}{9\sqrt{3}}K$\\ [0.1 cm]
&$^2{\bf 1}$&0&$\frac{1}{6\sqrt{3}}(2+K)$\\ [0.1 cm]
\hline
&total&$\frac{1}{6\sqrt{3}}(2+K)$&$\frac{1}{2\sqrt{3}}(2+K)$\\ [0.3 cm]
\hline
\hline
$\Xi ^0 \rightarrow \Lambda \gamma $
&$^2{\bf 8}$
&$-\frac{1}{6\sqrt{3}}(2+\frac{K}{3})$&$-\frac{1}{3\sqrt{3}}(2+\frac{K}{3})$
\\ [0.1 cm]
&$^4{\bf 8}$&$-\frac{1}{9\sqrt{3}}K$    &$-\frac{2}{9\sqrt{3}}K$\\ [0.1 cm]
&$^2{\bf 1}$&$\frac{1}{6\sqrt{3}}(2+K)$ &$0$\\ [0.1 cm]
\hline
&total&0 & $-\frac{1}{3\sqrt{3}}(2+K)$\\ [0.3 cm]
\hline
\hline
$\Xi ^0 \rightarrow \Sigma ^0 \gamma $ 
&$^2{\bf 8}$ &$\frac{1}{6}(2+\frac{K}{3})$&0
\\ [0.1 cm]
&$^4{\bf 8}$ &$\frac{1}{9}K$&0\\ [0.1 cm]
&$^2{\bf 1}$ &$\frac{1}{6}(2+K)$&$0$\\ [0.1 cm]
\hline
&total&$\frac{1}{3}(2+K)$&$0$\\ [0.3 cm]
\hline
\hline
\end{tabular}
\end{center}

\renewcommand{\baselinestretch}{1.4}

The weights given in Table 1 have to be multiplied by appropriate pole
factors.  In an idealized $SU(3)$-symmetric case, 
when all states of a given $SU(6)\times O(3)$ multiplet 
have the same mass, 
these factors are equal for $(b1)$ and $(b2)$
diagrams.
One can then see from Table 1 that Hara's theorem is satisfied
when the parity-violating WRHD amplitudes are proportional to the
differences of weights appropriate for diagrams $(b1)$ and $(b2)$.
This proportionality to weight differences is indeed obtained when
the relevant calculations are performed in a Hara's-theorem-satisfying
framework, as it was done in ref.\cite{Orsay}.

Omission in ref.\cite{Holstein99}
of the contribution from the singlet intermediate state
may affect asymmetries of the neutral hyperons only.  
Thus, only the estimates of the
parity-violating amplitudes of 
 $\Lambda \rightarrow n \gamma$, 
 $\Xi ^0 \rightarrow \Lambda \gamma$, and 
 $\Xi ^0 \rightarrow \Sigma ^0 \gamma $ performed in ref.\cite{Holstein99}
 should be done anew.
 On the basis of Eqs.(18) of ref.\cite{Holstein99} we may form a table
 of octet contributions arising there from diagrams $(b1)$ and $(b2)$.

Table 2

Weights of amplitudes corresponding to intermediate octet states in diagrams
$(b1)$ and $(b2)$ in ref.\cite{Holstein99}. Simplified formulas obtained
for $\omega _f=-\omega _d$ and corresponding directly to Table 1 are given
for each decay in bottom rows.

\begin{center}
\begin{tabular}{|l|c|c|c|}
\hline
process &  parameters  & diagram $(b1)$ & diagram $(b2)$ \\ [0.3 cm]
\hline
$\Lambda \rightarrow n \gamma $          
& $\omega _d,~\omega _f$ 
& $\frac{1}{6\sqrt{3}}(\omega _d +3\omega _f)$
&$-\frac{1}{6\sqrt{3}}(\omega _d -3\omega _f)$
\\ [0.1 cm]
&$\omega _f = - \omega _d$
&$\frac{1}{6\sqrt{3}}\cdot 2\omega _f$
&$\frac{1}{3\sqrt{3}}\cdot 2\omega _f$
\\ [0.2 cm]
\hline
\hline
$\Xi ^0 \rightarrow \Lambda \gamma $     
&$\omega _d,~\omega _f$ 
&$-\frac{1}{6\sqrt{3}}(\omega _d +3\omega _f)$
&$\frac{1}{6\sqrt{3}}(\omega _d -3\omega _f)$
\\ [0.1 cm]
&$\omega _f = - \omega _d$
&$-\frac{1}{6\sqrt{3}}\cdot 2\omega _f$
&$-\frac{1}{3\sqrt{3}}\cdot 2\omega _f$
\\ [0.2 cm]
\hline
\hline
$\Xi ^0 \rightarrow \Sigma ^0 \gamma $   
&$\omega _d,~\omega _f$
&$-\frac{1}{6}(\omega _d - \omega _f)$
&$\frac{1}{6}(\omega _d + \omega _f)$
\\ [0.1 cm]
&$\omega _f=-\omega _d$
&$\frac{1}{6}\cdot 2\omega _f$
&$0$
\\ [0.2 cm]
\hline
\hline
\end{tabular}
\end{center}

We rewrite Eqs.(18) of ref.\cite{Holstein99} in the form of Table 2, 
in notation analogous
to that used in Table 1.  
In Table 2 the pole factors are omitted and the
normalization is adjusted 
so as to simplify comparison of Tables 1 and 2.
(In ref.\cite{LZ,Orsay,Holstein99}, conventions of relative phases
between the 
$\Lambda \rightarrow n \gamma$, 
 $\Xi ^0 \rightarrow \Lambda \gamma$, and 
 $\Xi ^0 \rightarrow \Sigma ^0 \gamma $ amplitudes are identical.)

The fit used in ref.\cite{Holstein99} is characterized by
$\omega _f \approx 2370\cdot 10^{-7}~MeV$, 
$\omega _d \approx - 1780 \cdot 10^{-7}~MeV$ i.e. by
$
\omega _f/\omega _d \approx - 1.3 
$.
Proximity of the latter ratio to $-1$ is necessary if a reasonable description
of the experimentally very small $\Xi ^- \rightarrow \Sigma ^- \gamma $ 
branching
ratio is to be achieved.
Indeed, in ref.\cite{Holstein99}
the parity-violating amplitude of the
$\Xi ^- \rightarrow \Sigma ^- \gamma $ is smaller 
than the parity-violating amplitudes for the
decays of neutral hyperons by a factor of the order of 
$(\omega _f + \omega _d) \cdot (m_{\Xi }-m_{\Sigma })
/(m_{\Xi (\Sigma )}-m_{1/2^-})$.
The value of $\omega _f/\omega _d$ is equal to $-1$ if only
$W-$exchange processes contribute: 
the decay $\Xi ^- \rightarrow \Sigma ^- \gamma $,
being wholly due to a single quark transition, cannot then occur.
This was the assumption
made in ref.\cite{Orsay}. 
Consequently, in order to compare ref.\cite{Holstein99} with the original paper
\cite{Orsay}, we have to set $\omega _f = - \omega _d$ 
as it is done in respective
rows in Table 2.
One can see that the pattern of  the $^2{\bf 8}$ and $^4{\bf 8}$ 
contributions to the parity-violating amplitudes of neutral hyperon decays in 
ref.\cite{Orsay} (Table 1)
is identical to that in the BH paper for $\omega _d = -\omega _f$  
with the correspondence
$2+\frac{K}{3} \leftrightarrow 2\omega _f $ for $^2{\bf 8}$ 
(or $\frac{2}{3}K \leftrightarrow 2\omega _f$ for $^4{\bf 8}$).
In the fit in refs.\cite{LZ,Zen91}, 
apart from the contribution of $W-$exchange
processes determined (without any free parameters) 
from nonleptonic hyperon decays, the contribution from
single quark processes responsible for $\omega _d \neq - \omega _f$
was taken into account (and described by fit parameter) as well.

Since in ref.\cite{Orsay} $K \approx 1.25$, it follows 
from Table 1 that the weights of
contributions from $^4{\bf 8}$ resonances are smaller by a factor of around
$3$ than those due to $^2{\bf 8}$.  
In addition, the $^4{\bf 8}$ resonances are heavier 
than the $^2{\bf 8}$ resonances and the corresponding
pole factors should be smaller. Thus, in the first approximation
one might neglect the contribution of $^4{\bf 8}$ states, 
as it was done in ref.\cite{Holstein99}.
However, in ref.\cite{Orsay} 
the weights of singlet contributions are (for $K$ of the order of $1$)
of the same size as (or somewhat larger than)
those of the lowest-lying octet (Table 1).  
Since the singlet $\Lambda (1405)$ has the lowest mass of
all the excited $1/2^-$ resonances, 
its contribution is important
and has to be taken into account.

\section{Estimates of parity-violating amplitudes and of asymmetries}

Below we estimate the sign and size of the  $\Lambda (1405)$ contribution 
to the parity-violating 
$\Xi ^0 \rightarrow \Lambda \gamma $ amplitude, relative to
the sign and size of the octet contribution 
calculated in ref.\cite{Holstein99}.
In ref.\cite{Holstein99} the contributions from the
$\Lambda (1670)$ and $\Xi ^0 (?)$ combine to give the total parity-violating
amplitude (in units of $10^{-7}~GeV^{-1}$
used in ref.\cite{Holstein99}) as a sum
of contributions from diagrams $(b1)$ and $(b2)$:
\begin{eqnarray}
\label{HolXi0L}
B^{\Lambda \Xi^0} & = &
8\sqrt{2}~ er_d 
\left(\frac{1}{m_{\Xi}-m_{R_1}}\frac{1}{6\sqrt{3}}(\omega _d + 3 \omega _f)
+\frac{1}{m_{\Lambda }-m_{R_2}}\frac{1}{6\sqrt{3}}(\omega _d - 3 \omega _f)
\right)
\nonumber \\
&= &-0.58 + 0.50 = -0.08
\end{eqnarray}
where $e~r_d \approx 0.022~GeV^{-1}$
and resonance masses $m_{R_1} \approx m_{R_2} \approx 1540~MeV$ 
were used.
The first (second) term above comes from excited $J^P=1/2^-$ octet $\Lambda $ 
($\Xi ^0$) respectively.  In reality, both of these resonances are heavier 
than the value of $1540~MeV$ employed in \cite{Holstein99}, 
and one may expect their
contributions to $B^{\Lambda \Xi^0}$  to be somewhat smaller.
The near cancellation of the two contributions in Eq.(\ref{HolXi0L})
occurs also if more realistic masses of the resonances are used.
Thus, for $m_{R_1}=1670 ~MeV$ and $m_{R_2}=1830~MeV$ one gets
$B^{\Lambda \Xi^0}=-0.36+0.30=-0.06$.

In Table 1 
the contribution from the singlet is similar in absolute size to that 
from the octet $\Lambda $, but of opposite sign.
Thus, when the lowest-lying singlet and octet states 
are both taken into account, we expect that 
the parity-violating $B^{\Lambda \Xi^0}$ amplitude should be around
\begin{equation}
\label{Holcorr}
B^{\Lambda \Xi^0} \approx -0.58 +0.50 +0.4 \approx +0.3
\end{equation}
with the third term in the sum in Eq.(\ref{Holcorr}) resulting from 
a very rough 
(assuming  identical pole factors) estimate of the singlet contribution:
from Table 1 it should be close to $+0.58$ 
 if comparison with the $^2{\bf 8}$ $ \Lambda $  
 weight  (column $(b1)$) is employed
 or around $0.50/2=0.25$ if comparison with the $^2{\bf 8}$ $\Xi ^0$
  weight (column $(b2)$) is made.
 In Eq.(\ref{Holcorr}) the average of these two estimates is used.
 Similar estimates are obtained if one first uses 
 $\omega_f \approx -\omega_d \approx 
2075\cdot 10^{-7} ~MeV$ (the average of values used in \cite{Holstein99})
and then estimates 
the singlet contribution on the basis of
Table 1.
Using $m_{\Lambda _1}=1520~MeV$ 
one then obtains for the singlet contribution
\begin{equation}
\label{singlet}
\Delta B^{\Lambda \Xi ^0}(\Lambda _1) \approx \frac{1}{m_{\Lambda _1}-
m_{\Xi }}\cdot er_d~ \frac{8\sqrt{2}}{3\sqrt{3}}~\omega _f \approx +0.5
\end{equation}
instead of the value of $+0.4$ in Eq.(\ref{Holcorr}).

I think that putting the quark-model mass for $m_{\Lambda _1}$ 
(equal to $m(\Lambda (3/2^-))
\approx 1520~ MeV$) instead of the real mass of $\Lambda (1405)$,
is more appropriate here.
Explanation of the small $\Lambda (1405)$ mass is presumably connected
with coupling to $\bar{K}N$ and related channels.
If we were to use the value of $1405~MeV$ for the mass of $\Lambda _1$, we
would also have to include the effects of hadron-level corrections
to the size of weak and electromagnetic
hadronic transition amplitudes used in the model.
At present there is no model which could estimate such corrections
in a reliable way.

One may also give an experiment-based 
argument that in WRHD's the contribution from $\Lambda (1405)$ should
follow symmetry predictions given by the weights of Table 1, with all
pole factors of approximately the same size.
Namely, one may 
look at data on hyperon nonleptonic decays and,
assuming the dominance of $(70,1^-)$ contributions,
try to learn from the data about the properties of the 
contribution from the singlet $\Lambda $.
It turns out (compare Table 2 in ref.\cite{Orsay0})
that the excited singlet
$\Lambda $ contributes to $\Sigma ^+_+$ and $\Sigma ^-_-$
parity-violating
transitions, but not at all to those of $\Lambda$ or $\Xi $ decays.
For $\Sigma $ decays, 
the weights of contributions from the singlet ($^2{\bf 1}$)
and the lighter octet ($^2{\bf 8}$) were determined in ref.\cite{Orsay0}
to be of roughly the same size (up to a factor of  $1.5$ for $\Sigma ^-_-$,
and about $-0.5$ for $\Sigma ^+_+$).
As a result, if the size of the singlet contribution relative to 
that of the octet were modified too much by a completely different pole factor,
we should not be able to describe the $\Sigma$ decay amplitudes with
the same parameters that may be extracted from
the s-wave amplitudes of $\Lambda $ and $\Xi $.

Indeed, using  only the $\Lambda $ and $\Xi $ s-wave amplitudes, one can
extract $f-d \approx -2.83$, $ f+d \approx -0.91 $ 
(in units of $10^{-7}$).
As can be checked in Table 2 of ref.\cite{Orsay0},
the dominant contribution in these decays comes from the heavier
($^4{\bf 8}$) intermediate states (the ratio of weights from
$^4{\bf 8}$ and $^2{\bf 8}$ is $8:1$).
Assuming (for justification see below) 
that the contributions from the $^2{\bf 8}$ and $^4{\bf 8}$
intermediate states are given by symmetry considerations (i.e.
disregarding possible difference of the size of the relevant 
pole factors) and
denoting the contributions from both octets and the singlet 
by $O$ and $S$ respectively,
we then have for the $\Sigma $ decays (from \cite{Orsay0}):
\begin{eqnarray}
\Sigma ^+_+ = +0.13 = & O - S      & =  0                       \nonumber \\
\Sigma ^+_0 = -3.27 = & 3O/\sqrt{2}& =  \sqrt{\frac{3}{2}}(f-d) \nonumber \\
\Sigma ^-_- = +4.27 = & -2O -S     & = -\sqrt{3} (f-d)
\end{eqnarray}
where the column of numbers represents the data, 
and the rightmost entries give standard
expressions for the amplitudes in terms of $f$ and $d$.
The choice $S=O$ corresponds then to the symmetry situation in which
the contribution from the $\Lambda (1405) $ 
is evaluated with the pole factor
identical to that used for octet contributions.
Using $f-d$ determined from $\Lambda $ and $\Xi $ decays
one then predicts that
$\Sigma ^+_0 = -3.47$, and $\Sigma ^-_- =+4.90$, which is
in fair agreement with the data. 
(An overall fit to $\Lambda$, $\Xi $ {\em and} $\Sigma $ decays 
gives slightly different values for $f$ and $d$ and describes the data
a little better. 
Also, one has to remember that the
$\Delta I = 3/2$ contributions are of the order of a few percent, thus
defining what is the acceptable accuracy.)

Let us note now 
that the $\Sigma ^+_0 $ amplitude depends on intermediate octets only.
In fact, it follows from Table 2 of ref.\cite{Orsay0}
that the contribution to $\Sigma ^+_0 $ comes entirely from
the $^2{\bf 8}$ intermediate states.
The approximate equality $-3.47 \approx -3.27$ confirms therefore
that the relative size of 
contributions from $^2{\bf 8}$ and $^4{\bf 8}$ is properly
given by symmetry considerations (as assumed above), 
without taking into account
the difference in the size of pole factors.
The agreement $+4.90 \approx +4.27$ for the $\Sigma ^-_- $
amplitude indicates the same for the contribution from the singlet.
Had we used the singlet contribution enhanced 
by (say)  30\% (or 80\% ) due to a larger pole factor, 
we would have obtained $\Sigma ^-_- = +5.39~({\rm or}~ +6.21) $ respectively, 
in much worse agreement with the data. 
At the same time we would have obtained $\Sigma ^+_+ =
+0.49~({\rm or}~+1.30)$.
Clearly, in nonleptonic hyperon decays the singlet contribution follows
the symmetry prescription.
I think that this is a good hint that in WRHD's  the relative size of 
the contribution from the singlet $\Lambda $, relative to that from the octet
$\Lambda $, should follow the pattern of weights in Table 1.

With $\Delta B^{\Lambda \Xi ^0}(\Lambda _1)$ as in Eq.(\ref{singlet}),
one obtains
\begin{equation}
\label{otherwise}
B^{\Lambda \Xi^0}=-0.36+0.30 +0.5=+0.44
\end{equation}
in good agreement with Eq.(\ref{Holcorr}).

Using the value of the parity-conserving amplitude $A^{\Lambda \Xi^0}=-0.34$
from the BH paper, one obtains the asymmetry:
\begin{equation}
\alpha (\Xi ^0 \rightarrow \Lambda \gamma) \approx  ~-1.
\end{equation}
(in ref. \cite{Holstein99} this asymmetry was calculated to be $+0.46$),
and the decay rate (in units of GeV):
\begin{equation}
\Gamma (\Xi ^0 \rightarrow \Lambda \gamma) \approx (5~{\rm to}~6) \cdot 10^{-18} 
\end{equation}
($2.5\cdot 10^{-18}$ in the BH paper \cite{Holstein99} and in the experiment).
Let us note that in ref.\cite{Holstein99} the contribution of 
the excited $1/2^+$ states
to the $\Xi ^0 \rightarrow \Lambda \gamma $
parity-conserving amplitude was found to be negligible. 
In reality, this contribution is probably even smaller:
the size of the $\Sigma ^+ \rightarrow p \gamma $ branching ratio 
seems to indicate that
the overall size
of the contribution from excited $1/2^+$ states has been overestimated in BH. 
Consequently, the parity-conserving 
$\Xi ^0 \rightarrow \Lambda \gamma $ amplitude
is well estimated by the ground-state contribution.

The present analysis of the BH approach, 
with the $SU(3)$-singlet baryon contribution  
taken into account, shows that in Hara's-theorem-satisfying chiral framework
the $\Xi ^0 \rightarrow \Lambda \gamma $ asymmetry is large and negative,
in complete agreement with previous studies \cite{LZ,Orsay}.
Note also that if the weights $(b1)$ and $(b2)$ are added 
(as predicted by calculations in
Hara's-theorem-violating approaches \cite{LZ}), one gets 
\begin{equation}
B^{\Lambda \Xi^0} =
-0.58 - 0.50 + 0.40 \approx  -0.7
\end{equation} 
(alternatively: $-0.36-0.30+0.5 = -0.16$)
leading to a large and positive $\Xi ^0 \rightarrow \Lambda \gamma $ 
asymmetry (around +0.8).

For the $\Lambda \rightarrow n \gamma $ decay, 
Table 1 shows that in Hara's-theorem-satisfying 
approach of BH one obtains
$B^{n\Lambda} \approx 0.30-0.35-0.25 \approx -0.3$ 
(the first two figures in this sum come from ref.\cite{Holstein99},
the third one, i.e. $-0.25 \approx -0.3 \approx -0.35/2$  
is a rough estimate of the singlet contribution, obtained in a way
analogous to that for $\Xi ^0 \rightarrow \Lambda \gamma $). 
Since in ref.\cite{Holstein99}
the parity-conserving amplitude $A^{n\Lambda}$ 
is positive and equal to $+0.52$
(with contribution from excited $1/2^+$ states amounting to 25\% only), 
one concludes that the asymmetry should be large and negative.
Thus, when the singlet is taken into account,
the near-zero negative asymmetry of BH remains negative
but becomes much larger.
On the other hand, if amplitudes corresponding to diagrams 
$(b1)$ and $(b2)$ are added
(as in Hara's-theorem-violating cases), one gets
$B^{n\Lambda} \approx 0.30+0.35+0.25 \approx 0.9$ 
and a large positive asymmetry should be
observed.

For the $\Xi ^0 \rightarrow \Sigma ^0 \gamma $ decay,
Table 1 shows that with singlet contribution included,
the parity-violating amplitude will be of approximately twice the value
given in ref.\cite{Holstein99}, i.e. $B^{\Sigma ^0 \Xi ^0}\approx +1.4$.
Within the BH approach, 
this leads to a very small positive asymmetry (around $+0.07$).
Should the contribution of the excited $1/2^+$ states be smaller
than in ref.\cite{Holstein99}, 
the parity-conserving amplitude and
the resulting asymmetry
would become negative, irrespectively of whether the 
excited $J^P=1/2^-$ singlet is or 
is not taken into account.
This prediction of negative  $\Xi ^0 \rightarrow \Sigma ^0 \gamma $
asymmetry agrees nicely with the recent experiment,
according to which $\alpha (\Xi ^0 \rightarrow \Sigma ^0 \gamma) =
 -0.65 \pm 0.13$  \cite{Ramberg99,Koch99},
and is another indication that the contribution of the excited
$1/2^+$ resonances is overestimated in BH.
Because for $\Xi ^0 \rightarrow \Sigma ^0 \gamma $
there is almost no contribution from diagram $(b2)$ 
(in the original BH paper this 
contribution is negligible, while
in ref.\cite{Orsay} it is zero), 
the total $\Xi ^0 \rightarrow \Sigma ^0 \gamma $ amplitude
for Hara's-theorem-satisfying case is almost the same
as for Hara's-theorem-violating case.
Consequently, measurement of the  
asymmetry of the $\Xi ^0 \rightarrow \Sigma ^0 \gamma $ decay {\em alone}
does not provide useful information on the question of 
the violation of Hara's theorem.

\section{Summary}

In this paper we have analysed an extended version of the chiral model
of WRHD's discussed recently by Borasoy and Holstein \cite{Holstein99}.
In this version the contribution from the intermediate singlet baryon
has been properly taken into account.
The analysis of the signs of the
$\Xi ^0 \rightarrow \Lambda \gamma $,
$\Xi ^0 \rightarrow \Sigma ^0 \gamma $, and
$\Lambda \rightarrow n \gamma $ asymmetries
is then in complete accord with the discussion given previously in \cite{LZ}.
Of course, one has to remember that predictions of ref.\cite{Holstein99} are
based on a fit to nonleptonic hyperon decays obtained without 
taking into account the usual SU(3)-singlet classification of  
$\Lambda (1405)$. Consequently, 
the original fit should in principle 
be redone with the singlet included, and only then
WRHD's should be considered. 
It might seem that the discussion of the present paper would be meaningful
only provided such an improved fit had been done first.
Fortunately, this is not the case. 
Experimental smallness of the $\Xi ^- \rightarrow \Sigma ^- \gamma$
branching ratio proves that $\omega _f \approx - \omega _d$, irrespectively
of model details. Our analysis is based on this assumption 
(which is also
approximately satisfied in BH),
and on the relative smallness of contributions from excited
$1/2^+$ states in $\Xi ^0 \rightarrow \Lambda \gamma$ and 
$\Lambda \rightarrow n \gamma $ decays.

In summary, in the chiral approach of Borasoy and Holstein, asymmetries of the
$\Xi ^0 \rightarrow \Lambda \gamma $ and $\Lambda \rightarrow n \gamma $ decays 
are both large and negative
when the $SU(3)$ singlet assignment of $\Lambda (1405)$ is taken into account.
This should be contrasted with Hara's-theorem-violating approaches in which
these asymmetries are large and positive.

\section{Acknowledgments}
I would like to thank B. Holstein for a discussion concerning the issue of
(un?)assailability of fundamental assumptions needed to prove Hara's theorem.


\end{document}